# Non-axisymmetric flow characteristics in Head-on Collision of Spinning Droplets

Chengming He (何成明) and Peng Zhang* (张鹏)

*Department of Mechanical Engineering, The Hong Kong Polytechnic University, Hung Hom, Kowloon, Hong Kong, 999077*

Effects of spinning motion on the bouncing and coalescence between a spinning droplet and a non-spinning droplet undergoing the head-on collision were numerically studied by using a Volume-of-Fluid method. A prominent discovery is that the spinning droplet can induce significant non-axisymmetric flow features for the head-on collision of equal-size droplets composed of the same liquid. Specifically, a non-axisymmetric bouncing was observed, and it is caused by the conversion of the spinning angular momentum into the orbital angular momentum. This process is accompanied by the rotational kinetic energy loss due to the interaction between the rotational and radial flows of the droplets. A non-axisymmetric internal flow and a delayed separation after temporary coalescence were also observed, and they are caused by the enhanced interface oscillation and internal-flow-induced viscous dissipation. The spinning motion can also promote the mass interminglement of droplets because the locally non-uniform mass exchange occurs at the early collision stage by non-axisymmetric flow and is further stretched along the filament at later collision stages. In addition, it is found that the non-axisymmetric flow features increase with increasing the orthogonality of the initial translational motion and the spinning motion of droplets.

---

* Author to whom correspondence should be addressed.
E-mail address: pengzhang.zhang@polyu.edu.hk (P. Zhang)

# NOMENCLATURE

*Physical quantities*

$D$      Droplet diameter

$M$      Mass of the droplet

$t$      Physical time

$t_{osc}$      Characteristic oscillation time, $t_{osc} = \sqrt{\rho_l D_l^3 / \sigma_l}$

$U$      Experimental relative velocity between two colliding droplets

$\mu$      Dynamic viscosity

$\rho$      Density

$\sigma$      Surface tension coefficient

$\chi$      Projection of two mass centers connection line in the direction perpendicular to $U$

$\Omega$      Dimensional angular velocity

*Non-dimensional and normalized variables*

$B$      Impact parameter, $B = \chi / D_l$

$Oh$      Ohnesorge number, $Oh = \mu_l / \sqrt{\rho_l D_l \sigma_l}$

$T$      Non-dimensional time, $T = t / t_{osc}$

$We$      Weber number, $We = \rho_l D_l U^2 / \sigma_l$

$\boldsymbol{\omega}$      Angular velocity vector

*Mathematical and numerical parameters*

| | |
|---|---|
| $A$ | Colored contact surface |
| $c$ | Volume fraction |
| $H(c-1)$ | Heaviside step function limits the integration domain to be within the droplets |
| $I$ | Moment of inertia |
| $\boldsymbol{L}$ | Angular momentum vector |
| $N$ | Mesh refinement level / the number of meshes |
| $V$ | Integral volume of liquid and gas phases |
| $\boldsymbol{v}$ | Velocity vector varying in space |
| $\boldsymbol{V}$ | Velocity vector given at specific positions |
| $\boldsymbol{r}$ | Position vector varying in space |
| $\boldsymbol{R}$ | Position vector given at specific positions |
| $\boldsymbol{i}, \boldsymbol{j}, \boldsymbol{k}$ | Unit vectors in the $x$-, $y$-, and $z$- directions of the mass center coordinate system |
| $\theta$ | Polar angle between the spinning axis $l_{O_1}$ and $z$-axis |
| $\varphi$ | Azimuthal angle between the projection of spinning axis $l_{O_1}$ on the $x$-$y$ plane and $x$-axis |
| $\phi$ | Mass dye (color) function |
| $\delta_s$ | Dirac delta function |
| $\kappa$ | Local curvature |

*Subscripts*

| | |
|---|---|
| 0 | Values at the initial instant |
| 1 | Properties of droplet $O_1$ |
| 2 | Properties of droplet $O_2$ |
| c | Values of characteristic scale |
| g | Properties of the surrounding gas phase |
| l | Properties of the liquid droplet |
| o | Orbital component of angular momentum |
| s | Spinning component of angular momentum |
| t | Total angular momentum |

## I. INTRODUCTION

Collision of two liquid droplets in gaseous environment is ubiquitous in nature and industries. Many experimental studies have been reported[1-10] and reviewed[11-13] in the literature. Most of them were focused on identifying and interpreting various outcomes of droplet collision [3-5, 7-9], such as coalescence, bouncing, separation and shattering[14, 15], rendering a well-known collision nomogram in the $We - B$ parameter space. The collision Weber number, $We$, which measures the relative importance of the droplet inertia compared to the surface tension, and the impact parameter, $B$, which measures the deviation of the trajectory of droplets from that of the head-on collision, with $B = 0$ denoting the head-on collision and $B = 1$ the grazing collision. Besides, the influences of some other controlling parameters on the collision outcomes have also been investigated, such as the droplet Ohnesorge number[7, 9, 16, 17], $Oh$, which measures the relative importance of the liquid viscous stress compared to the capillary pressure, and the size ratio[3, 18-20], $\Delta$, which measures the droplet size disparity. In addition, the collision outcomes can be significantly affected by the gas environment as such increasing the gas pressure promotes droplet bouncing and decreasing the gas pressure promotes droplet coalescence[4, 5]. A practical significance of the gas pressure effect is that colliding fuel droplets may tend to bounce off under high pressure in real combustion chambers, and it has been verified both experimentally and numerically[21-23].

It has been recognized that the colliding droplets usually have a spinning motion. The spinning motion can be created either from droplet injectors (by nonuniform driving pressure) or from preceding collisions, which are off-center as a large probability. Bradley and Stow[2] showed the experimental images of droplet spin after coalescence and measured the angle of rotation as a function of time and impact parameter. Ashgriz and Poo[3] proposed the schematic of reflexive separation for the off-center droplet collision by considering the droplet spin after coalescence.

Rotational energy[4, 6] has been considered in various models for predicting outcomes of off-center droplet collisions. The fact that the spinning motion of a droplet after off-center collisions can take part of energy from its translational motion has however not been considered in the previous models. It is noted that the spinning droplets can collide with each other because subsequent collisions are highly probable in practical dense sprays[21, 24, 25], but relevant studies have not been seen in the literature.

The spinning effects on single droplet have been investigated by a number of studies. Brown and Scriven[26] used a finite-element method to trace the equilibrium state of axisymmetric, two-, three- and four-lobed drop shape for rotating droplets and analyzed the critical transition between different drop shapes. Kitahata *et al.*[27] proposed a simple mechanical model to measure the liquid surface tension by use of the frequency-amplitude relation of oscillation of a levitated rotating droplet. Holgate and Coppins[28] numerically studied the equilibrium shapes and stability of rotating charged drops in a vacuum and proposed a formula for stability limit. The phenomena of droplet distortion and spinning can be further enriched by the involvement of rotating environmental flows[29, 30] or external fields[31].

In this paper, we shall present a computational study on the collision of spinning droplets. This study restricts its scope to the head-on collision between droplets of equal size so as to avoid unnecessary complexity of geometrical asymmetry and size disparity, which certainly merit future studies. Furthermore, the ideally spherical shape of droplets is assumed before their collision, because the characteristic rotational energy, $M_l \Omega^2 D_l^2 / 4$, due to the droplet spinning motion is substantially smaller than the surface energy, $\pi \sigma_l D_l^2$, with the ratio $M_l \Omega^2 / 4\pi \sigma_l$ being $O(10^{-1})$ or less (see the following section for the estimation of angular velocity $\Omega$). Consequently, the spinning droplet is negligibly deformed due to its centrifugal force. The presentation of the study is organized as follows. The numerical methodology and specifications are described in Sec. II. The results of the bouncing and coalescence between a spinning droplet and a non-spinning droplet are presented in Sec. III and IV, respectively.

## II. NUMERICAL METHODOLOGY AND SPECIFICATIONS

### A. Numerical method

The three-dimensional continuity and incompressible Navier-Stokes equations,

$$\nabla \cdot \boldsymbol{v} = 0 \tag{1}$$

$$\rho(\partial \boldsymbol{v}/\partial t + \boldsymbol{v} \cdot \nabla \boldsymbol{v}) = -\nabla p + \nabla \cdot (2\mu \boldsymbol{D}) + \sigma \kappa \boldsymbol{n} \delta_s \tag{2}$$

are solved by using the classic fractional-step projection method, where $\boldsymbol{v}$ is the velocity vector, $\rho$ the density, $p$ the pressure, $\mu$ the dynamic viscosity, and $\boldsymbol{D}$ the deformation tensor defined as $D_{ij} = (\partial_j u_i + \partial_i u_j)/2$. In the surface tension term $\sigma \kappa \boldsymbol{n} \delta_s$, $\delta_s$ is a Dirac delta function, $\sigma$ the surface tension coefficient, $\kappa$ the local curvature, and the unit vector $\boldsymbol{n}$ normal to the local interface.

To solve both the gas and liquid phases, the density and viscosity are constructed by the volume fraction as $\rho = c\rho_l + (1-c)\rho_g$ and $\mu = c\mu_l + (1-c)\mu_g$, in which the subscripts $l$ and $g$ denote the liquid and gas phases, respectively. The volume fraction $c$ satisfies the advection equation

$$\partial c/\partial t + \nabla \cdot (c\boldsymbol{v}) = 0 \tag{3}$$

with $c = 1$ for the liquid phase, $c = 0$ for the gas phase, and $0 < c < 1$ for the gas-liquid interface. The present study simulating droplet collisions adopts the Volume-of-Fluid (VOF) method, which has been implemented in the open source code, Gerris[32, 33], featuring the three-dimensional octree adaptive mesh refinement, the geometrical VOF interface reconstruction, and continuum surface force with height function curvature estimation. Gerris has been demonstrated to be competent for solving a wide range of multiphase flow problems[19, 34-40].

A major challenge of VOF simulations on droplet collision lies in the inability of the Navier-Stokes equations in describing the rarified gas effects and the Van der Waals force[41] within the gas film, thereby prohibiting the physically correct prediction of droplet coalescence. A coarse mesh would induce "premature" coalescence of the droplets that realistically bounce off. Thus, the successful simulation of droplet coalescence and subsequent collision dynamics in previous studies[19, 20, 34-36] were obtained by choosing an appropriate mesh resolution near the interface. Similarly, the conventional VOF simulation on droplet bouncing requires an extremely refined mesh and a substantial computational cost, we thereby adopted two VOF functions[37, 38, 42] to separately track the interface of each liquid droplet so as to avoid interface coalescence on a relatively coarse mesh. The method with two VOF functions was successfully applied to study droplet bouncing and has been verified by experiments. It can produce nearly the same droplet deformation and minimum interface distance for droplet bouncing cases compared to the conventional VOF approach. Furthermore, the droplet interfaces can be advected in the two immediately neighboring interface cells, leading to the minimum interface distance can be smaller than the minimum mesh size. More details about the two VOF methods and their comparison have been sufficiently discussed in the literature[37, 38, 42].

## B. Numerical validations

To validate the present numerical method, the head-on droplet bouncing and coalescence were simulated and compared with the experimental results of Pan *et al.*[10] and Qian and Law[5], respectively. To improve computational efficiency, the computational domain is divided into three physical zones, namely the gas, the droplet, and the interface, and each zone has its own mesh refinement level denoted by $N$, which corresponds to a minimum mesh size of $O(2^{-N})$. Accordingly, $(N_g, N_l, N_i)$ is used to describe the refinement level in the three zones. As a balance between computational cost and accuracy in the present study, a mesh refinement level of (3, 5, 7) was used (by He et al.[37]) for all simulations on droplet bouncing by using the approach of two VOF functions, and a mesh refinement level of (4,7,8) was used (by Chen et al.[34]) for all simulations on droplet coalescence based on the conventional VOF method.

The numerical validations with experiments and grid independence analysis of droplet bouncing and coalescence were respectively conducted in our previous studies[37] and Chen et al.[34], in which the same simulation on droplet coalescence at $We = 61.4$ and $Oh = 2.80 \times 10^{-2}$ ($t_{osc} = 1.06ms$) was reproduced and compared with the experimental images from Qian and Law[5] in Fig. 1. The experimental and simulation times are nearly identical in early collision stages and begin to display slight discrepancies as time evolves in later stages. The time errors are generally less than 8%. The discrepancies between experiment and simulation may be attributed to both possible numerical errors and unideal experimental conditions. Specifically, the experimental images of droplet deformation show apparently asymmetric features, implying small errors in the experimental setup and control, but the droplets are always perfectly spherical and identical in the numerical setup.

Given the maximum mesh refinement level $N_i$ in the interface zone, the maximum numerical resolution (MNR)[38] of a droplet can be defined as $MNR = 2^{N_i} + 1$. A typical simulation run with the mesh refinement level (3, 5, 7) results in $MNR = 129$ and 240,700 grid points in the entire droplet, which is equivalent to about $2.0 \times 10^8$ grid points if applying a uniform mesh with size of $O(2^{-7})$. It takes about 100 hours of real time to run the simulation up to T = 2.0 on an Intel Xeon(R) E5-2630 processor with 16 cores. Similarly, a typical simulation run with the mesh refinement level (4, 7, 8) results in MNR = 257 and $3.73 \times 10^6$ grid points in the entire droplet, taking about 200 hours of real time to run the simulation up to $T = 2.0$ on two Intel Xeon(R) Gold-6150 processor with 72 cores (36 cores for each processor).

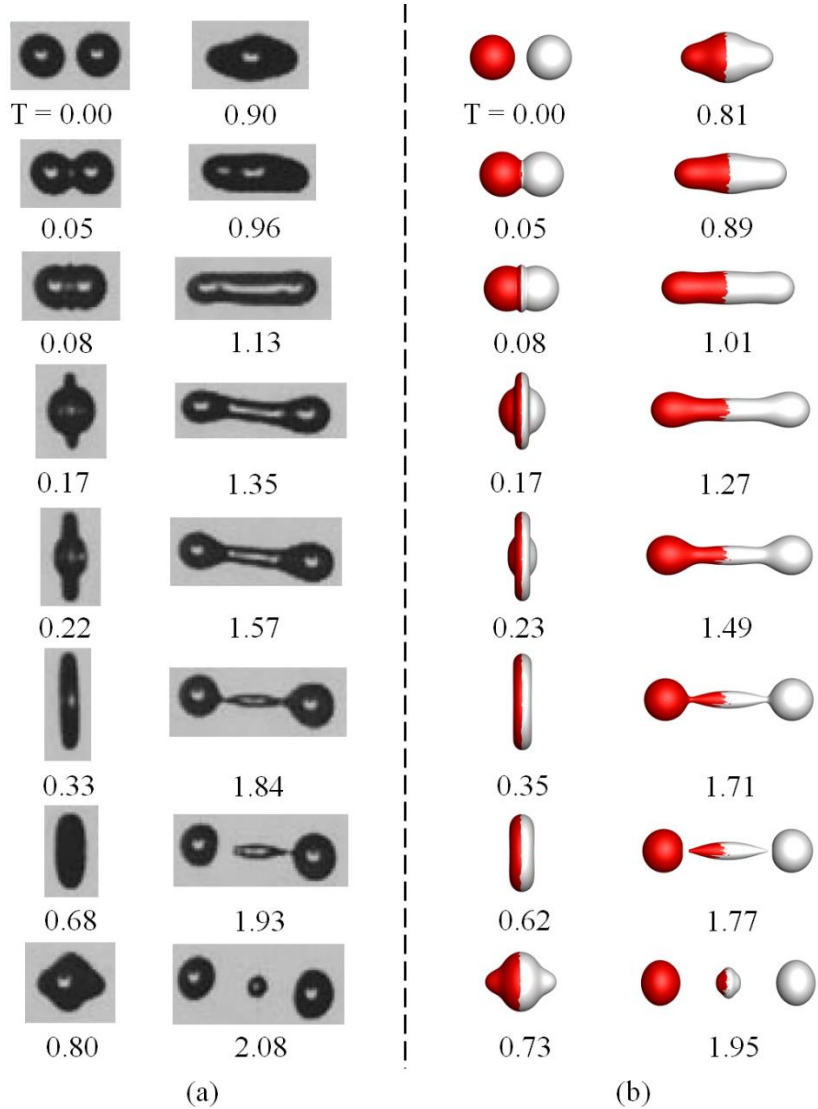

FIG. 1. Comparison of (a) experimental images adapted from Qian and Law[5] and (b) the present numerical results for the head-on coalescence of identical droplets at $We = 61.4$ and $Oh = 2.80 \times 10^{-2}$. The dimensionless time $T = t/t_{osc}$ and $t_{osc} = 1.06$ ms.

## C. Evaluation of droplet spinning speed

A key process in setting up the initial conditions for the present simulations is to specify the physically realistic spinning speed of droplets. It has not been seen in the literature on the experimental measurement of the spinning speed of droplets after collisions, probably because of the difficulty of experimentally resolving the droplet spinning motion under the sub-mini scales of length and time. Consequently, we analyzed the simulation data in the literature[37] to evaluate the physically realistic range of the rotational speed for a droplet that is made to spin as the result of a preceding collision.

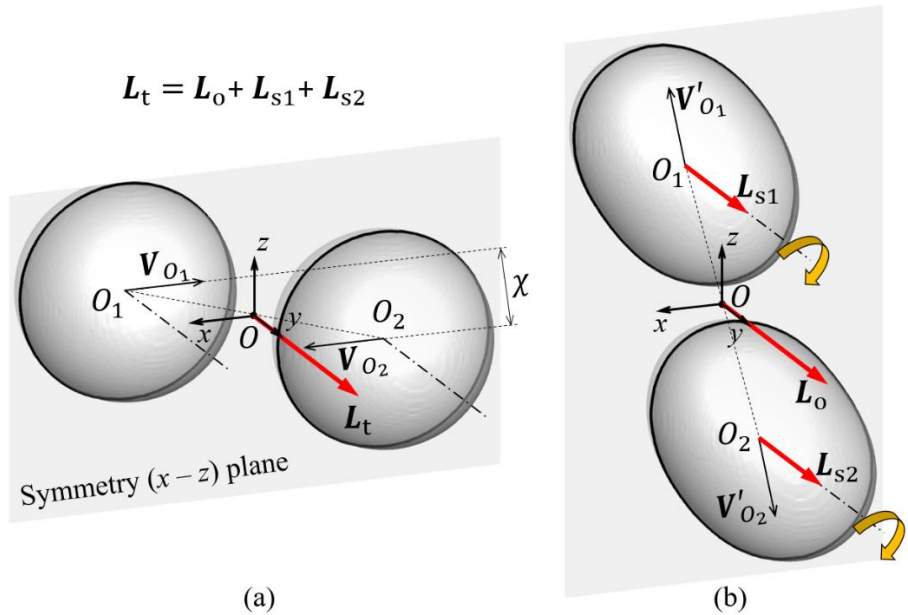

FIG. 2. Schematic of arbitrary off-center droplet collision involving a symmetry (*x-z*) plane and conservation of angular momentum (a) before and (b) after the collision, in which $L_t$, $L_o$, $L_{s1}$ and $L_{s2}$ are the total, orbital, and spinning (subscript "1" and "2" denote two droplets) angular momentum and have only one component in the *y*-direction.

Figure 2 shows the general schematic of an off-center droplet collision. There always exists a symmetry (*x-z*) plane[37, 43] that is established by the *x*-axis and the connection line $O_1O_2$ (hereinafter referred to $\overline{O_1O_2}$) of the mass centers of the colliding droplets. As the evolution of droplet deformation, the origin of the mass center coordinate system is always located at the midpoint of $\overline{O_1O_2}$ owing to the vanishing linear momentum in the coordinate system, and the velocity vectors before and after the collision are denoted as ***V*** and ***V*′**. Subscripts "$O_1$" and "$O_2$" are denoted for two droplets at each mass center, respectively. The velocity vectors are always on the symmetry plane for the head-on collision, which is a special case with the relative velocity being parallel to $\overline{O_1O_2}$.

The angular momentum ***L*$_o$** and ***L*$_s$** are shown in Fig. 2, where ***L*$_o$** is the orbital angular momentum with respect to the *y*-axis, and ***L*$_s$** for each liquid droplet is the spin angular momentum with respect to the spinning axis across its mass center. As the evolution of droplet deformation, the position vectors of the mass centers $O_1$ and $O_2$ for two droplets can be numerically calculated by

$$\boldsymbol{R}_{O_1} = \int_V \rho H(c_1 - 1)\boldsymbol{r}\, dV / M_1 \tag{4a}$$

$$\boldsymbol{R}_{O_2} = \int_V \rho H(c_2 - 1)\boldsymbol{r}\, dV / M_2 \tag{4b}$$

Consequently, ***L*$_o$** and ***L*$_s$** can be further expressed by

$$\boldsymbol{L}_o = \int_V \left[\rho H(c_1 - 1)\boldsymbol{R}_{O_1} + \rho H(c_2 - 1)\boldsymbol{R}_{O_2}\right] \times \boldsymbol{V}\, dV \tag{5}$$

and

$$\boldsymbol{L}_{s1} = \int_V \rho H(c_1 - 1)(\boldsymbol{r} - \boldsymbol{R}_{O_1}) \times \boldsymbol{V}\, dV \tag{6a}$$

$$\boldsymbol{L}_{s2} = \int_V \rho H(c_2 - 1)(\boldsymbol{r} - \boldsymbol{R}_{O_2}) \times \boldsymbol{V}\, dV \qquad (6b)$$

Apparently, $\boldsymbol{L}_t$, $\boldsymbol{L}_o$ and $\boldsymbol{L}_s$ vanish for head-on collisions. Whereas the off-center droplet collision has a symmetry (*x-z*) plane, the angular momentums have only one component in the *y*-direction. Consequently, based on the numerical calculation of $L_{s1}\boldsymbol{j}$ (same to $L_{s2}\boldsymbol{j}$ owing to the symmetry), the angular velocity of droplet 1 or 2 (hereinafter referred to D1 or D2) in *y*-direction can be estimated by $\omega \boldsymbol{j} = L_{s1}\boldsymbol{j}/I_1$, where the inertia of moment $I_1$ of D1 is a scalar that numerically calculated by

$$I_1 = \int_V \rho H(c_1 - 1)(\boldsymbol{r} - \boldsymbol{R}_{O_1})^2\, dV \qquad (7)$$

The rotational deformation and spinning angular velocity of droplets are characterized in Fig. 3 by two representative cases of off-center droplet bouncing at $We = 9.3$ and $We = 20$, which have been discussed in detail in our previous paper[37]. As shown in Fig. 3(a), the solid point and line denote the time-dependent center of mass and the initial intersecting plane, respectively. The non-dimensional angular velocity $\omega$ in Fig. 3(b) increases rapidly at the early stage from T=0.0 to about 0.5 and then remains unchanged during late stages. Droplet rotation after off-center collisions favors larger $We$, with $\omega \approx 1.25$ for $We = 9.3$ and $\omega \approx 2.0$ for $We = 20$, which corresponds to dimensional angular velocities $\Omega = \omega \Omega_c$ of 1886 rad/s and 1179 rad/s, respectively. Here, $\Omega_c$ is the characteristic angular velocity given by $\Omega_c = 1/t_{osc} = 943$ rad/s. Furthermore, the collision of droplets in elevated pressure environment[4, 5] tends to bounce at a higher $We$ and thereby induce a stronger droplet spinning. Thus, the present study adopts a range of $\omega = 1.0 \sim 3.0$, which corresponds to $\Omega$ being about $1000 \sim 3000$ rad/s.

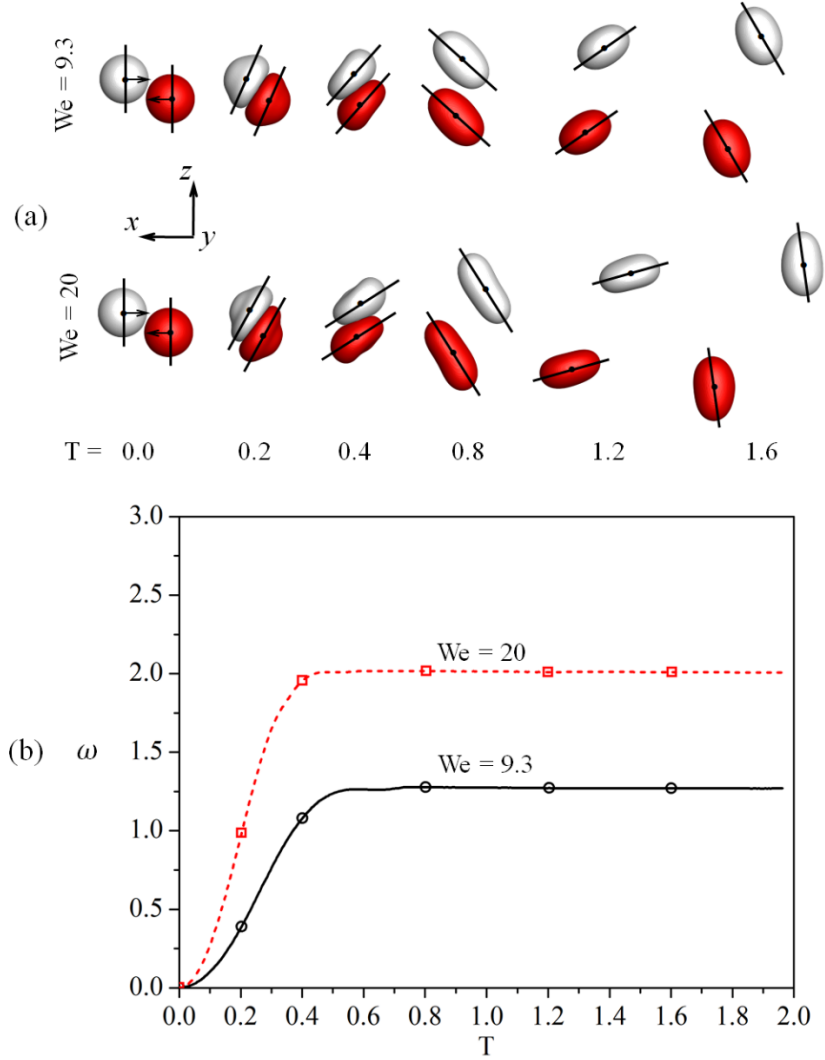

FIG. 3. Evolution of (a) droplet deformation and spinning motion and (b) non-dimensional angular velocity $\omega$ for bouncing droplets at $B = 0.4$ and $Oh = 2.8 \times 10^{-2}$ at $We = 9.3$ and $We = 20$.

## D. Problem description and numerical specifications

The 3D computational domain of the head-on collision (with vanishing impact parameter $B$) between a spinning D1 and a non-spinning D2 is illustrated in Fig. 4. The spinning axis $l_{O_1}$ of D1 can be described by a polar angle $\theta$ with respect to the $z$-axis and an azimuthal angle $\varphi$ to the $x$-axis. In the present study, the polar angle $\theta$ is fixed at $\pi/2$ and the azimuthal angle varies in the range of $0 \leq \varphi \leq \pi/2$. Then, the initial spinning angular velocity can be expressed as $\boldsymbol{\omega}_0 = (-\omega_0\cos\varphi, -\omega_0\sin\varphi, 0)$.

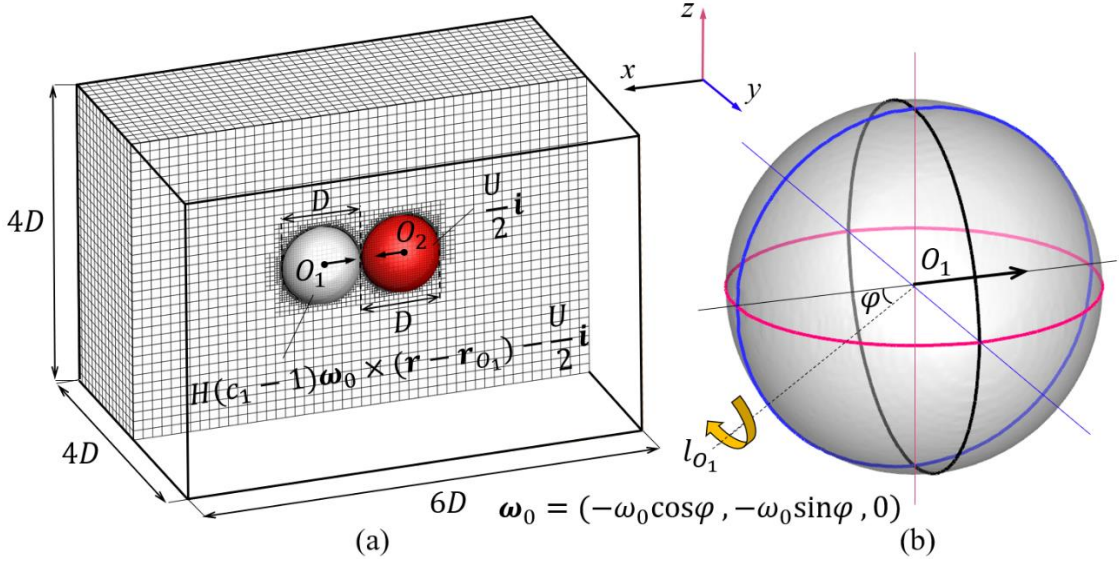

FIG. 4. Schematic of (a) three-dimensional computational domain and (b) setup of spinning axis $l_{O_1}$ and initial velocity vector for the head-on collision between a spinning droplet 1 and a non-spinning droplet 2.

Two droplets of diameter $D$ are specified to collide along the $x$-direction with a relative translational velocity, $U$, and therefore they have zero relatively velocities in the $y$- and $z$-directions. The translational velocity component for D1 and D2 are set as $-\frac{U}{2}\boldsymbol{i}$ and $\frac{U}{2}\boldsymbol{i}$, respectively, so that the linear momentum of the entire mass-center system remains zero. The spinning velocity components of D1 is given by $H(c_1 - 1)\boldsymbol{\omega}_0 \times (\boldsymbol{r} - \boldsymbol{R}_{O_1})$. The domain is $6D$ in length and $4D$ in both width and height with all boundaries specified with the free outflow boundary conditions. To avoid the unnecessary complexity and to emphasize on the spinning effects on the droplet collision dynamics, the present numerical study focuses on the representative case at fixed $Oh = 2.8 \times 10^{-2}$ and $\omega_0 = 3.0$.

## III. SPINNING-AFFECTED BOUNCING UPON HEAD-ON COLLISION

### A. Spinning-induced off-center bouncing ($\varphi = \pi/2$)

The head-on collision between a spinning D1 and a non-spinning D2 at $We = 9.3$ and $Oh = 2.8 \times 10^{-2}$ is shown in Fig. 5(b). The head-on collision between two non-spinning droplets are shown in Fig. 5(a) for comparison. The representative case with $\varphi = \pi/2$ has been concerned first, and thereby the spinning angular velocity for D1 is given by $\boldsymbol{\omega}_0 = -\omega_0 \boldsymbol{j}$. Due to the symmetry breaking by the spinning motion, three-dimensional flow features appear in the results, and they are illustrated on both the symmetry (*x-z*) plane and the $yO_1O_2$ plane, where the *y*-axis and $\overline{O_1O_2}$ lie, are shown in Fig. 5(b). For the non-spinning case, the axis-symmetric results are illustrated on the symmetry (*x-z*) plane, as shown in Fig. 5(a),

Some similarities are observed for these two cases in terms of the evolution of droplet deformation. The droplet interaction results in the locally enhanced capillary pressure around the rim of the interaction region where local curvature is large. The maximum droplet deformation is reached at about T = 0.35. The deformed droplets bounce back (at about T = 0.90) driven by the surface tension force, meanwhile converting the surface energy back to the kinetic energy. Subsequently, the droplets experience several oscillation periods before completely recovering their initial spherical shapes, which happens at later times beyond those shown in Fig. 5.

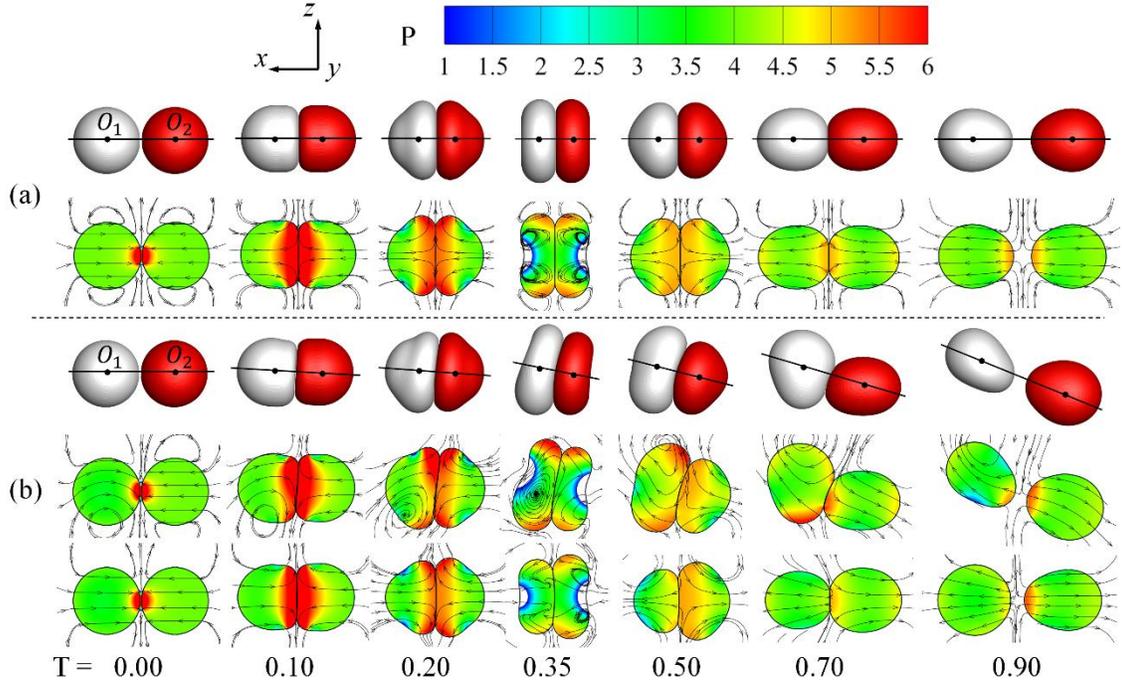

FIG. 5. Comparison of deformation, pressure profiles, and streamlines for the head-on droplet collision. (a) two non-spinning droplets and (b) a spinning droplet 1 ($\boldsymbol{\omega}_0 = -\omega_0 \boldsymbol{j}$) with a non-spinning droplet 2. The results are shown on the *x-z* plane only for cases (a) because they are axisymmetric, but both on the *x-z* plane at first two rows and on the plane $yO_1O_2$ consisting of *y* axis and $\overline{O_1O_2}$ at the last row for case (b).

Prominent difference can be seen for the two cases. Compared with the non-spinning case in Fig. 5(a), the non-axisymmetric droplet deformation with $\overline{O_1O_2}$ deviated from *x*-axis is observed in Fig. 5(b). This is attributed to the existence of interchanges between $\boldsymbol{L}_\text{t}$, $\boldsymbol{L}_\text{o}$, and $\boldsymbol{L}_\text{s}$, as shown in Fig. 6. Specifically, $\boldsymbol{L}_\text{s1}$ (initially equals to $\boldsymbol{L}_\text{t}$) causes a small increase of $\boldsymbol{L}_\text{s2}$ and a prominent increase of $\boldsymbol{L}_\text{o}$. It implies that the spinning motion of D1 can slightly rotate the non-spinning D2 but induces a prominent non-axisymmetric flow even by means of a head-on collision. For the present bouncing cases, the interchange of $\boldsymbol{L}_\text{s}$ and $\boldsymbol{L}_\text{o}$ can be realized only by the means of the intervening gas film between the droplets. To illustrate the role of the gas film, we show the vorticity amplitude |**ω**| and velocity vector in Fig. 7. As the closeup of gas film region shown in Fig. 7, the active shear flow in D1 by the initial spinning motion can induce the passive shear flow in D2 by the gas film, and consequently lead to the non-axisymmetric collision appearance with the mass center of droplets deviated off the *x*-axis.

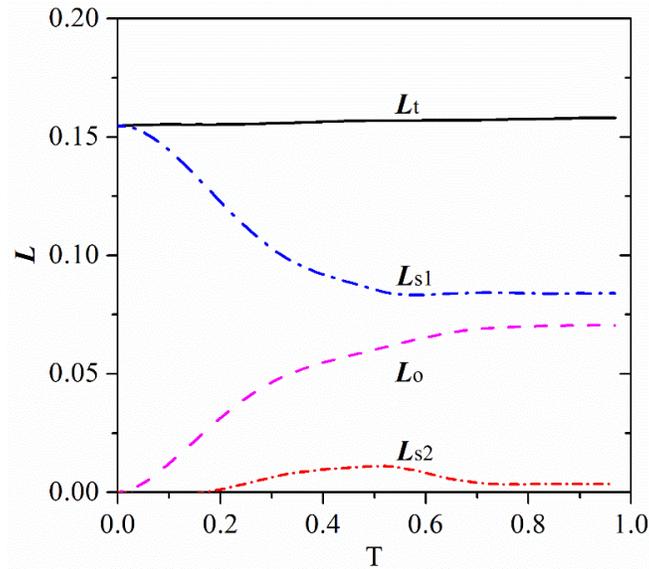

FIG. 6. Interchange between spinning angular momentum, $L_{s1}$ and $L_{s2}$, and orbital angular momentum, $L_o$, for the head-on collision between a spinning droplet 1 and a non-spinning droplet 2 shown in Fig. 5(b).

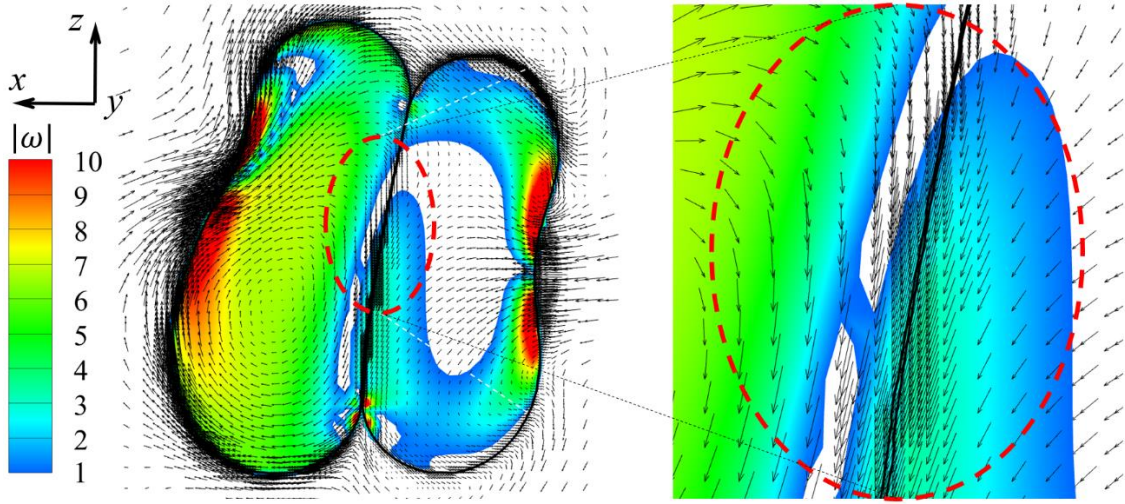

FIG. 7. Effects of gas film on the internal flow (vorticity magnitude $|\boldsymbol{\omega}|$ and velocity vector on the symmetry $x$-$z$ plane) by the spinning motion of droplets shown in Fig. 5(b) and Fig. 6. The white region inner droplets denote the $|\boldsymbol{\omega}|$ less than 1.0, which has been blanked for clear comparison of the vorticity concentration.

## B. Influence of the orientation of spinning axis ($\varphi \neq \pi/2$)

Apart from the special case with $\varphi = \pi/2$, the more general cases with varying spinning axes at $\varphi = 0, \pi/6, \pi/4$ and $\pi/3$ were studied to reveal the angular momentum interchange between two droplets. It is seen that $L_t$ in each direction is conserved at time instants of T = 0.0 and T = 1.0, as shown in Fig. 8, and only $L_{s1}$ presents at T = 0.0 because of the initial head-on collision.

Similar to the case of $\varphi = \pi/2$, a part of the initial $L_{s1}$ is transferred to $L_o$ and $L_{s2}$ to induce the spinning motion of D2 and orbital motion of two droplets. More specifically, for $L_x$ shown in Fig. 8(a), the initial $L_{s1}$ decreases with increasing $\varphi$, and very small $L_o$ and $L_{s2}$ are observed at T=1.0, implying that the interaction between D1 and D2 is weak on the $y$-$z$ plane and cannot cause the apparent non-axisymmetric flow or the spinning motion of D2. For $L_y$ shown in Fig. 8(b), the initial $L_{s1}$ increases with increasing $\varphi$; regardless of the negligible $L_{s2}$, a prominent $L_o$ after droplet collision is observed and reaching a maximum value at $\varphi = \pi/2$. For $L_z$ shown in Fig. 8(c), although the initial $L_{s1}$ is zero, nonzero $L_{s1}$, $L_o$, and $L_{s2}$ occur at T=1.0 at intermediate azimuthal angle, with $L_t$ is still conserved being zero. In summary, the most significant non-axisymmetric flow by the spinning effects occurs in the direction parallel to the *x-z* plane and would be enhanced as increasing $\varphi$. This finding implies that the orthogonality of the initial translational motion of two droplets and the spinning motion of D1 might be an important factor on generating the non-axisymmetric flow.

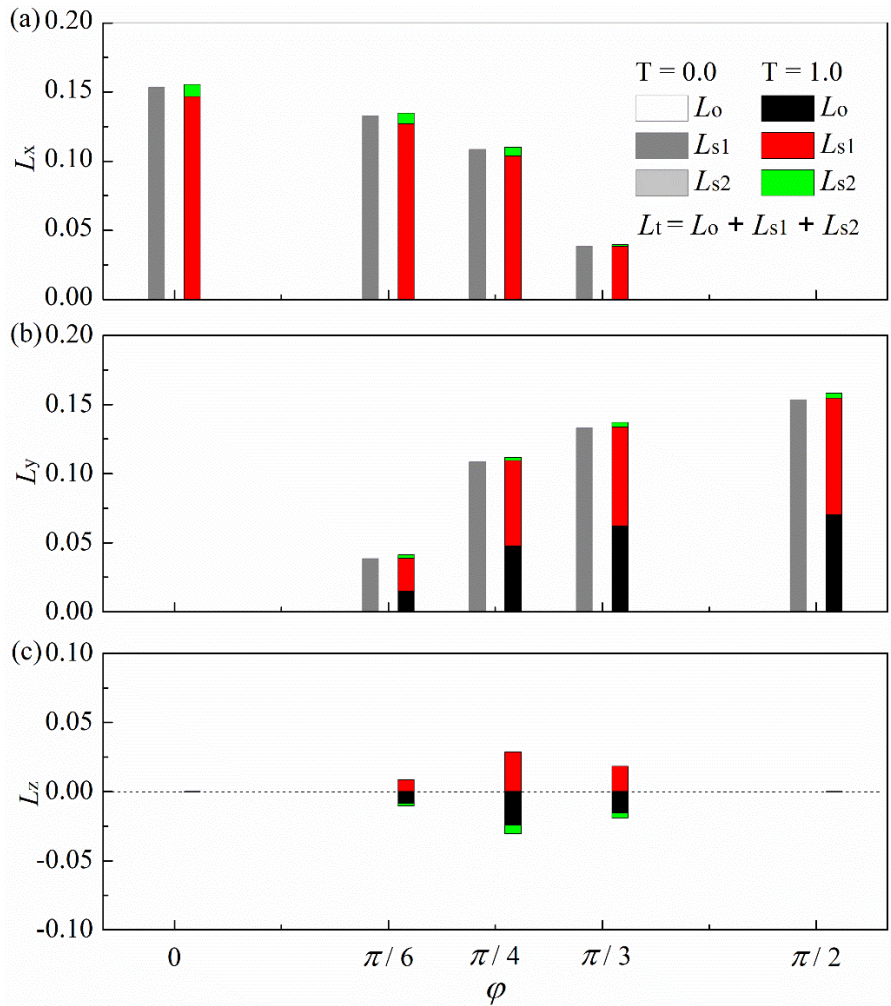

FIG. 8. Comparison of spinning angular momentum, $L_{s1}$ and $L_{s2}$, and orbital angular momentum, $L_o$, at time instants of T = 0.0 and T = 1.0 with varying the azimuthal angle $\varphi$. (a) $L_x$, (b) $L_y$, and (c) $L_z$ are the angular momentum in $x$-, $y$-, and $z$- directions, respectively.

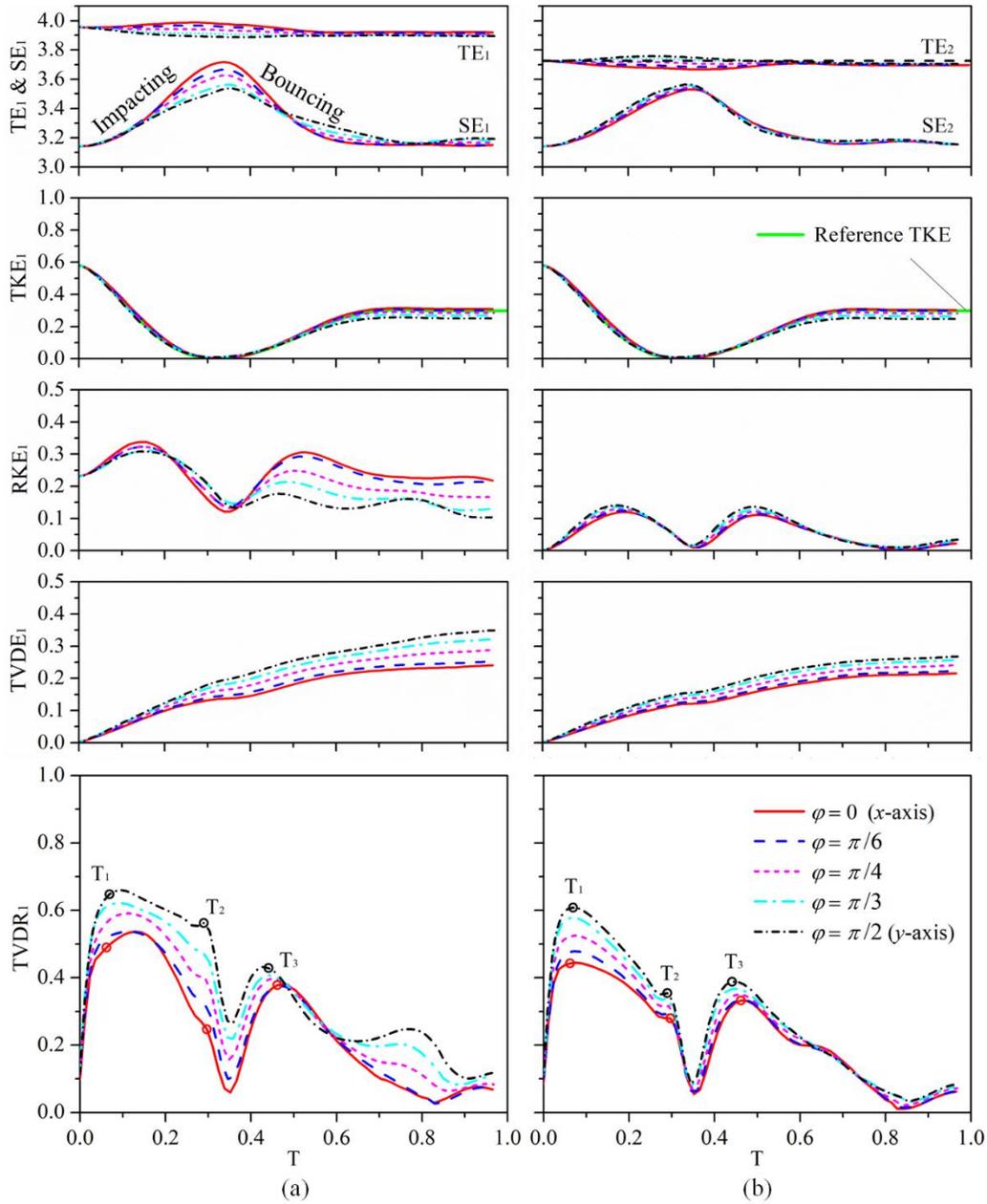

FIG. 9. Energy budget of head-on collisions between (a) a spinning droplet 1 and (b) a non-spinning droplet 2 with varying the azimuthal angle $\varphi$ shown in Fig. 8. The total energy (TE), the surface energy (SE), the translational kinetic energy (TKE), the rotational kinetic energy (RKE), the total viscous dissipation energy (TVDE), and the total viscous dissipation rate (TVDR) are nondimensionalized for each liquid droplet separately. The reference TKE is for the collision between two non-spinning droplets.

## C. Roles of rotational kinetic energy

Based on the extrema of surface energy (SE) curve, as shown in Fig. 9, the entire collision process can be divided into impacting, bouncing, and oscillating stages[37]. The impacting and bouncing stages are mainly concerned in the present study. The energy budget is analyzed for the spinning D1 and non-spinning D2 separately, with excluding the negligible energy in the gas phase[37]. The initial total energy ($TE_0$) of D1 is defined as $TE_0 = SE_0 + TKE_0 + RKE_0$, with $SE_0 = \pi$, the translational kinetic energy $TKE_0 = \pi We/48$, and the rotational kinetic energy $RKE_0 = \omega_0^2 \pi/120$. The initial total energy of D2 is $TE_0 = SE_0 + TKE_0$.

It is seen that the TE = SE + TKE + RKE + TVDE is approximately constant during the entire collision process, justifying again the neglect of the energy budget in the gas phase. Here, the total viscous dissipation energy (TVDE) is defined as

$$\text{TVDE}(T) = \int_0^T \left( \int_V H(c_1 + c_2 - 1) \phi \, dV \right) dT \tag{8}$$

where $\phi$ is the local viscous dissipation rate (VDR) given by

$$\phi = 2\mu \left[ \left(\frac{\partial u}{\partial x}\right)^2 + \left(\frac{\partial v}{\partial y}\right)^2 + \left(\frac{\partial w}{\partial z}\right)^2 \right] + \mu \left[ \left(\frac{\partial u}{\partial y} + \frac{\partial v}{\partial x}\right)^2 + \left(\frac{\partial v}{\partial z} + \frac{\partial w}{\partial y}\right)^2 + \left(\frac{\partial w}{\partial x} + \frac{\partial u}{\partial z}\right)^2 \right] \tag{9}$$

The separated $TE_1$ and $TE_2$ show that the energy is transferred from D2 to D1 with $\varphi$ smaller than $\pi/4$ while be transferred from D1 to D2 with $\varphi$ larger than $\pi/4$, The energy interchange between D1 and D2 must by means of the KE. Compared with the reference TKE in Fig. 9 of the head-on collision between two non-spinning droplets, the spinning effects of D1 have insignificant influences on the evolution of TKE, with the curves of TKE$_1$ and TKE$_2$ collapse with each other in the early stage and show slight differences in the later stage. It is thereby inferred that the RKE plays a dominant role on the energy interchange between D1 and D2. Specifically, RKE$_1$ and RKE$_2$ oscillate as the evolution of droplet deformation with changed inertia of moment and angular velocity. RKE$_2$ and RKE$_1$ with small $\varphi$ can recover to their initial value after several oscillations, whereas RKE$_1$ with large $\varphi$ shows the apparent loss to the viscous dissipation, as the TVDE$_1$ curve shown in Fig. 9(a).

The viscous dissipation varies monotonically with $\varphi$ for the representative case at $We = 9.3$ and $Oh = 2.8 \times 10^{-2}$. This can be understood by studying the interaction between the rotating flow parallel to the *x-z* plane owing to the spinning motion of D1 and the radial flow induced by the collision between D1 and D2. Specifically, as shown in Fig. 9(a) during the droplet impacting stage, with increasing $\varphi$ the TVDE$_1$ increases whereas the total increment of SE$_1$ and total decrement of KE$_1$ = TKE$_1$ + RKE$_1$ decrease, which is seemingly contradicted to the previous understanding that a larger droplet deformation with more KE converted into SE should invoke a larger TVDE due to the enhanced intensity of internal flow. However, in the axisymmetric internal flow at $\varphi = 0$, as shown in Fig. 10(b), apart from the inertia force responsible for the droplet radial spreading, the centrifugal force by the spinning motion of D1 is also in the radial direction and can further promote the radial spreading. So it cannot cause large velocity gradient accounting for the viscous dissipation. This is manifested by the enlarged droplet spreading diameter (L$_1$ > L$_2$) and nearly same VDR distribution for D1 and D2 shown in Fig. 10(b). Compared with the collision between two non-spinning droplets shown in Fig. 10(a), the spinning motion of D1 in Fig. 10(b) only slightly changes the droplet deformation and VDR distribution, which implies that the rotational energy serves as a "spring effect" and is not converted into the viscous dissipation owing to the weak interaction between rotating flow and radial flow.

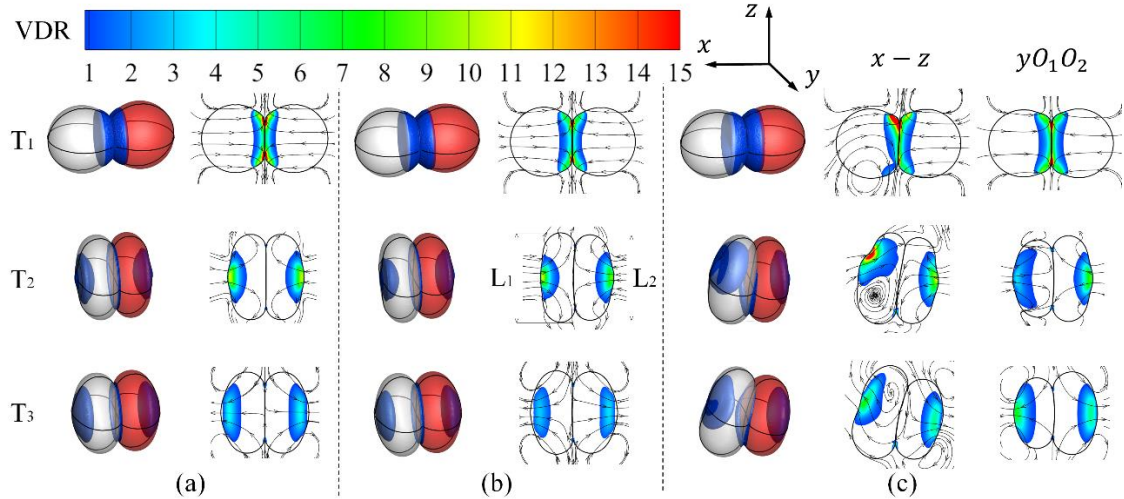

FIG. 10. Contour of the local viscous dissipation rate (VDR) and streamlines for the cases at three chosen time instants, $T_1$, $T_2$, and $T_3$ shown in Fig. 9, with (a) the collision between two non-spinning droplets at same time instants, (b) $\varphi = 0$, and (b) $\varphi = \pi/2$. The contours have been blanked with a low threshold value of 0.5 for clear comparison of the VDR concentration.

The substantial $TVDE_1$ at $\varphi = \pi/2$ can be explained as the strong interaction between the rotating flow and radial flow. As the VDR concentration on the *x-z* plane shown in Fig. 10(c) at time instant $T_1$, the spinning-induced rotating flow and the radial flow in D1 interact with each other to form an intense non-axisymmetric flow. This interaction causes enhanced velocity gradient and viscous dissipation at one side of the rim region of the spreading droplet. Whereas the internal flow of D1 and D2 on plane $yO_1O_2$ in Fig. 10(c) at time instant $T_1$ are symmetric and in radial direction. The interaction between the rotating flow and the radial flow presents mainly on the *x-z* plane at time instants of $T_2$ and $T_3$ and leads to the non-axisymmetric internal flow and enhanced viscous dissipation, as shown in Fig. 10(c). Consequently, we can conclude that the rotational kinetic energy loss is responsible for the non-axisymmetric flow by the interaction between the rotating and radial flows.

## IV. SPINNING-AFFECTED COALESCENCE UPON HEAD-ON COLLISION

### A. Non-axisymmetric droplet deformation and delayed separation

Figure 11 compares the results of the collision-induced coalescence between two non-spinning droplets and that between a spinning D1 (left) and a non-spinning D2 (right) at the representative case of $We = 61.4$ and $Oh = 2.8 \times 10^{-2}$. Only azimuth angle $\varphi = \pi/2$ is considered here because the preceding section has suggested that the interaction between the rotating flow and the radial flow tends to be maximal at $\varphi = \pi/2$. Furthermore, it should be emphasized that we aim to focus on studying the spinning effects on droplet coalescence by consistent numerical comparison through the same numerical setup and mesh parameters expatiated in Sec. II have been used. We do not intend to study the spinning effects on gas film drainage and on the interface rupture, which indeed merits future study but cannot be considered in the present computational framework.

It is seen that the merged droplet reaches its maximum deformation by depleting all kinetic energy, it contracts from a "disk" to a "filament", and finally leads to separation with a satellite droplet. Compared with the case of non-spinning droplets in Fig. 11(a), where the symmetric separation occurs at about T = 1.74, the non-symmetric separation for the spinning case has been delayed to about T = 1.81. Specifically, the liquid filament on the side of D2 is pinched off earlier than the other side of D1, leading to the formed satellite droplet being closer to D1. This is attributed to the non-axisymmetric flow induced by the spinning motion of D1, as shown by the streamline and pressure contour on the symmetry (*x-z*) plane in Fig. 11(b). The internal flow on the specific planes, which are parallel to the *y*-axis and denoted by a series of solid lines, still holds the mirror symmetry with respect to the *x-z* plane.

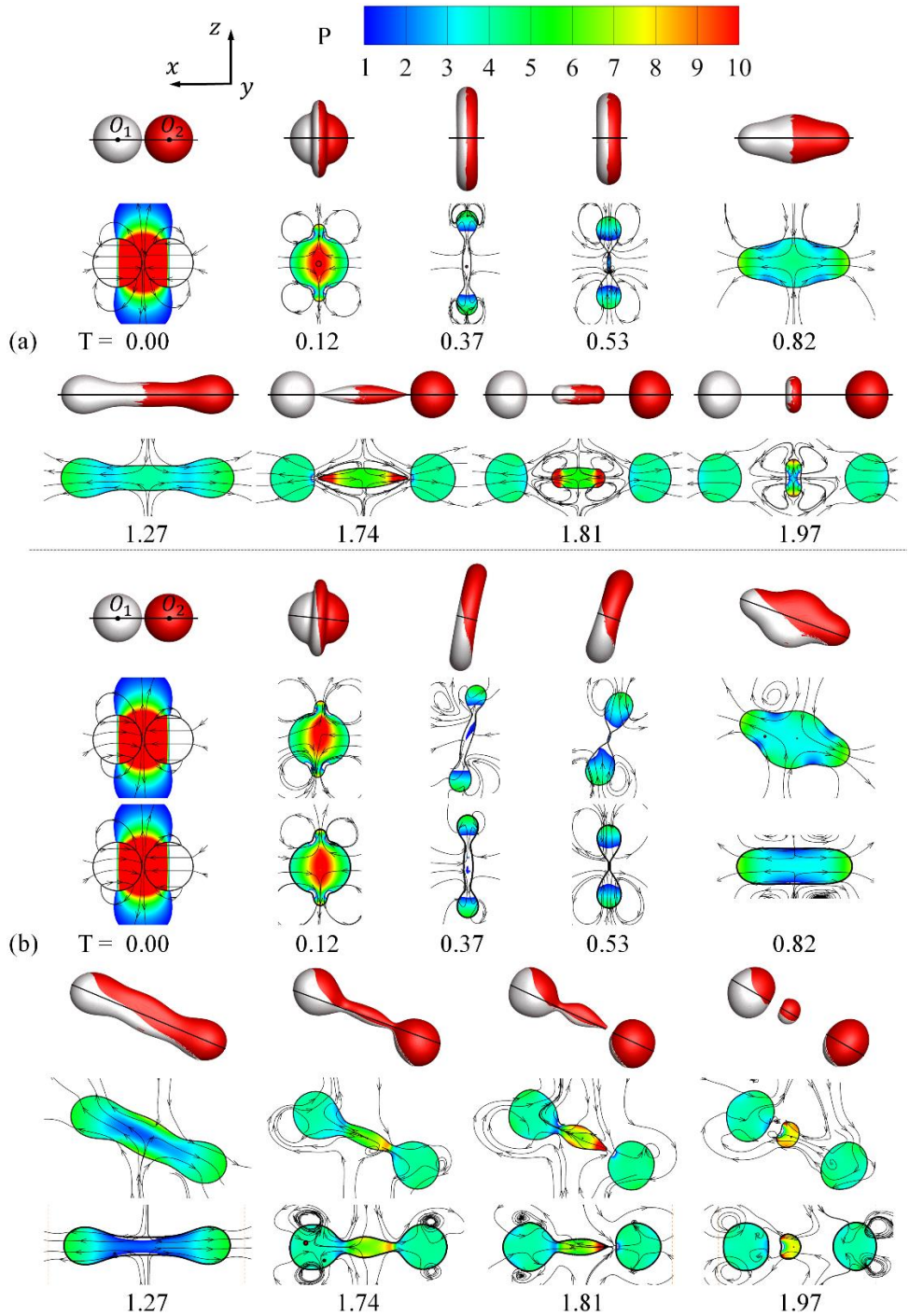

FIG. 11. Comparison of deformation, pressure profiles, and streamlines for the droplet coalescence and subsequent separation at $We = 61.4$ and $Oh = 2.8 \times 10^{-2}$. (a) two non-spinning droplets and (b) a spinning droplet 1 ($\boldsymbol{\omega}_0 = -\omega_0 \boldsymbol{j}$) with a non-spinning droplet 2. The third line of case (b) is the contour on the plane that parallel to the *y*-axis and illustrated by solid lines.

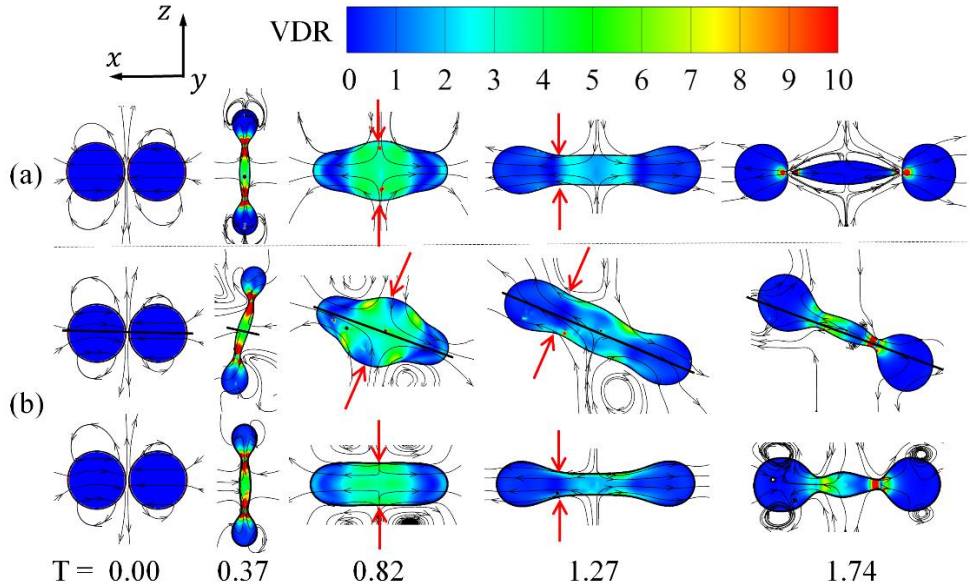

FIG. 12. Local viscous dissipation rate (VDR) and streamlines for the coalescence shown in Fig. 11. (a) two non-spinning droplets and (b) a spinning droplet 1 with a non-spinning droplet 2. The legend of VDR is divided by 5 for T = 0.82 and T = 1.27.

In the previous studies[4, 5, 18], a delayed or suppressed separation is usually attributed to the enhanced internal-flow-induced viscous dissipation with increasing liquid viscosity or droplet interminglement duration. The present observation of delayed separation is also caused by the enhanced viscous dissipation, but it is owing to the interaction between the rotating flow and the radial flow. This can be verified by the local VDR and streamlines shown in Fig. 12. Specifically, as the droplet contracts from a "disk" to a "filament", Fig. 12(a) and 12(b) show nearly the same filament length at T = 0.82. However, as shown by the red arrows, the case(b) has a larger filament width on the *x-z* plane while has a smaller filament width on the specific planes that denoted by a series of solid lines. The non-axisymmetric droplet deformation would promote the interface oscillation and cause enhanced internal-flow-induced viscous dissipation, as the enhanced VDR in the vicinity of the oscillating interface on the *x-z* plane shown at T = 1.27. Consequently, the viscous dissipation accompanied by the non-axisymmetric interface oscillation leads to the filament length of case (b) prominently shorter than that of case (a) at T = 1.27 and 1.74, so that it requires more time to separate.

**B. Enhanced internal mass interminglement by spinning effects**

It is also interesting to note that, in Fig. 11, regardless of the satellite droplet contain the liquid mass from both D1 and D2, mass exchange between D1 and D2 can be observed for the spinning case but not in the non-spinning case. It implies that the spinning effects can promote the internal mass interminglement by breaking the mirror symmetry of the collision between two identical droplets.

To quantitatively characterize the mass interminglement between D1 and D2, the temporal area changes of the colored contact surface of droplet were numerically calculated and shown in Fig. 13. The contact surface area $A(t)$ of the droplet is normalized by the initial surface area $A_0$ of the droplets. Because the mesh resolutions on the gas-liquid interface and inside the droplet are fixed in the present simulation setup, the normalized colored contact surface area $A(t)/A_0$ can be approximately calculated by[20]

$$\frac{A(t)}{A_0} = \frac{N[0 < \phi(t) < 1]H[c(t) - 1]}{N[0 < c(0) < 1]} \tag{10}$$

where $N$ is the number of the meshes in which the VOF function $c(t)$ or the mass dye (color) function $\phi(t)$ take certain values within their ranges, and thus it can be treated as functionals of $c(t)$ or $\phi(t)$. The color function $\phi(t) = 1$ denotes the spinning D1, $0 < \phi(t) < 1$ the contact surface of the droplets, and $\phi(t) = 0$ the non-spinning D2. Again, the Heaviside step function ensures only those meshes within the droplet are counted in the calculation. It should be noted that the mass interminglement concerned in the present study is not the physical mixing, and $A(t)/A_0$ is calculated by the numerical dyer function without liquid diffusion effects.

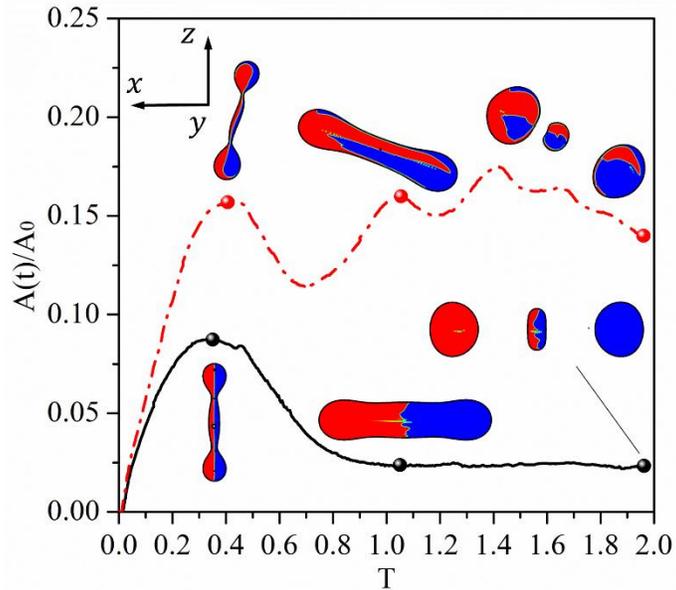

FIG. 13. Evolution of temporal contact surface area $A(t)/A_0$ to characterize the mass interminglement between the droplets shown in Fig. 11, with solid line for two non-spinning droplets and dashed dot line for a spinning droplet 1 with a non-spinning droplet 2.

Figure 13 shows the comparison between the evolution of $A(t)/A_0$ for the two coalescence cases shown in Fig. 11. The results show that the spinning motion of D1 can lead to prominent increase of $A(t)/A_0$, because the colored contact surface is stretched along the filament owing to the non-axisymmetric flow, as illustrated by the embedded internal mass interminglement on the *x*-*z* plane shown in Fig. 13.

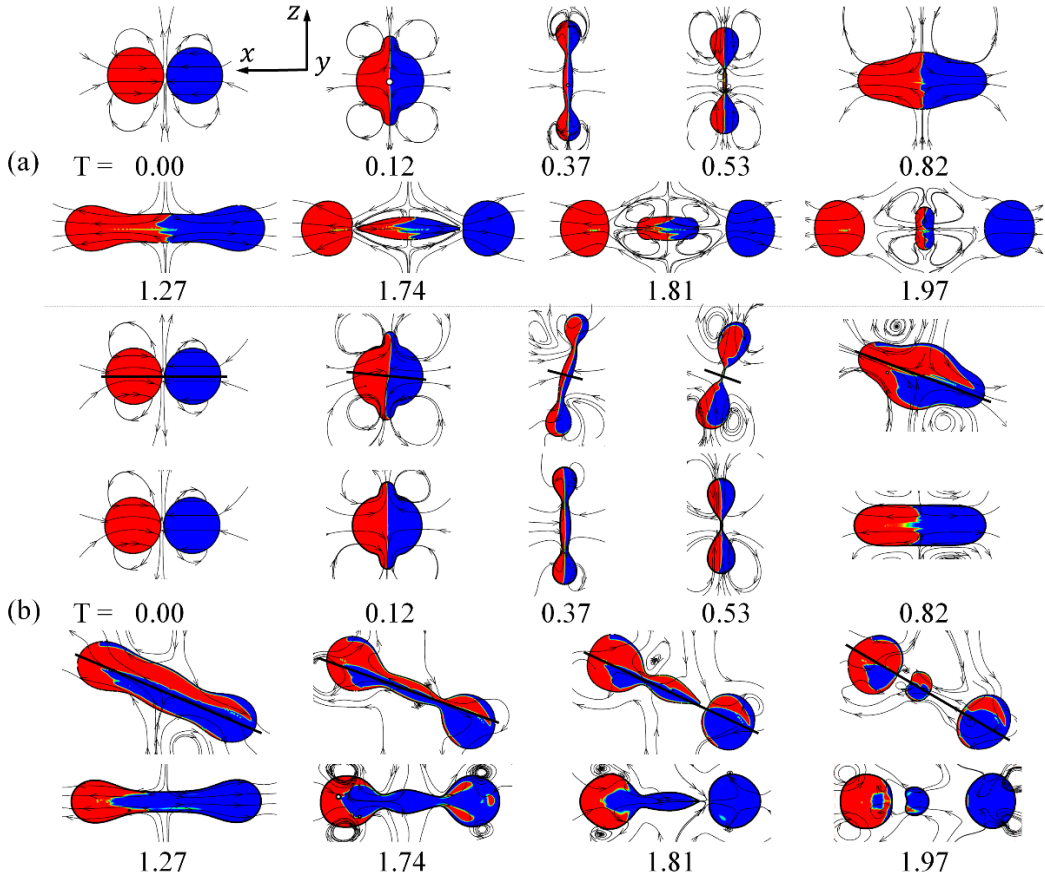

FIG. 14. Internal mass interminglement for (a) two non-spinning droplets and (b) a spinning droplet 1 (left) with a non-spinning droplet 2 (right) shown in Fig. 11.

Figure 14 shows the entire process of the enhanced mass interminglement and internal liquid mass stretching along the filament by the spinning effects. Specifically, as the mass interminglement shown at T = 0.12 and 0.37 in Fig. 14(b), the liquid mass from D1 by spinning effects would pass through the mirror symmetry plane of the head-on collision of equal-size droplets composed of the same liquid, and lead to locally non-uniform mass distribution around the rim of the interaction region. In the meantime, the colored contact surface inner the merged droplet is rotated and deviated from the *x-y* plane. Then, the merged droplet with non-uniform mass distribution begins to contract from the inclined "disk-like" deformation to an inclined filament, as the contour at T = 0.53 and 0.82 on the *x-z* plane shown in Fig. 14(b), and further being stretched along the filament in the subsequent droplet deformation. It is noted that the mass interminglement on the specific planes, which are parallel to the *y*-axis and denoted by a series of solid lines in Fig. 11, still holds the mirror symmetry with respect to the *x-z* plane. Consequently, it can be understood that the non-axisymmetric flow by the spinning effects at the early impacting stage would be enlarged in later droplet deformation stage and lead to the enhanced mass interminglement.

## V. CONCLUDING REMARKS

A computational study on the head-on bouncing and coalescence between a spinning droplet and a non-spinning droplet was investigated based on a validated Volume-of-Fluid method. The most interesting discovery is that the spinning motion has a significant role in affecting both droplet bouncing and coalescence.

For the head-on droplet bouncing, the spinning motion of droplet can induce non-axisymmetric flow features, because a part of the initial spinning angular momentum is converted into the orbital angular momentum. With varying the spinning axis of droplet, the non-axisymmetric flow becomes the most significant when the spinning axis is perpendicular to the direction of relative velocity. This indicates that the orthogonality of the initial translational motion of two droplets and the spinning motion of droplet is an important factor on enhancing the non-axisymmetric flow. In the aspects of energy conversion, the translational kinetic energy after droplet bouncing is not sensitive to the variation of spinning axis, whereas the apparent rotational kinetic energy loss to the viscous dissipation is attributed to the interaction between the rotating flow induced by droplet spinning motion and the radial flow induced by the droplets translational impacting motion.

For the head-on droplet coalescence, the spinning motion of droplet leads to a delayed separation after temporary coalescence, compared with the case between two non-spinning droplets. The delayed coalescence is attributed to the enhanced interface oscillation and internal-flow-induced viscous dissipation due to the non-axisymmetric droplet deformation. Furthermore, the spinning effects can significantly promote the mass interminglement by breaking the mirror symmetry of the head-on collision of equal-size droplets composed of the same liquid. This is because the non-axisymmetric flow by the spinning effects leads to locally non-uniform mass interminglement at the early collision stage, and because the contact interface between the mass from different droplets is further stretched along the filament in the later collision stages.

It should be noted that, although the present study deliberately limits its scope to the spinning effects on the collision between a spinning droplet and a non-spinning droplet, the discovered phenomena are believed to exist in general and may be more substantial in appropriate collision situations. The spinning effects on the droplet collision in practical situations are more complex than those investigated in the present study. For example, the collision of two spinning droplets necessitates the consideration of the relative orientation of two spinning axes and the spinning directions (clockwise or counter clockwise with respect to the axes). Apparently, the size disparity, the impact parameter, and the intensity of droplet spinning speed are of significance in practical situations and merited further investigations in the future. The last but not the least, the experimental confirmation of the present results may be of interest but certainly requires innovation of the current experimental techniques in generating and controlling spinning droplets.

**ACKNOWLEDGMENTS**

This work was supported partly by the Hong Kong RGC/GRF (Grant No. PolyU 152651/16E) and partly by the Hong Kong Polytechnic University (Grant Nos. G-SB1Q and G-YBXN).

**REFERENCE**

[1]. P. Brazier-Smith, S. Jennings, and J. Latham, The interaction of falling water drops: coalescence, Proc. R. Soc. Lond. A **326**, 393 (1972).

[2]. S. Bradley, and C. Stow, Collisions between liquid drops, Proc. R. Soc. Lond. A **287**, 635 (1978).

[3]. N. Ashgriz, and J. Poo, Coalescence and separation in binary collisions of liquid drops, J. Fluid Mech. **221**, 183 (1990).

[4]. Y. J. Jiang, A. Umemura, and C. K. Law, An Experimental Investigation on the Collision Behavior of Hydrocarbon Droplets, J. Fluid Mech. **234**, 171 (1992).

[5]. J. Qian, and C. K. Law, Regimes of coalescence and separation in droplet collision, J. Fluid Mech. **331**, 59 (1997).

[6]. J.-P. Estrade, H. Carentz, G. Lavergne, and Y. Biscos, Experimental investigation of dynamic binary collision of ethanol droplets–a model for droplet coalescence and bouncing, Int. J. Heat Fluid Flow **20**, 486 (1999).

[7]. M. Sommerfeld, and M. Kuschel, Modelling droplet collision outcomes for different substances and viscosities, Exp. Fluids **57**, 187 (2016).

[8]. G. Finotello, R. F. Kooiman, J. T. Padding, K. A. Buist, A. Jongsma, F. Innings, and J. Kuipers, The dynamics of milk droplet–droplet collisions, Exp. Fluids **59**, 17 (2018).

[9]. K. H. Al-Dirawi, and A. E. Bayly, A new model for the bouncing regime boundary in binary droplet collisions, Phys. Fluids **31**, 027105 (2019).

[10]. K.-L. Pan, C. K. Law, and B. Zhou, Experimental and mechanistic description of merging and bouncing in head-on binary droplet collision, J. Appl. Phys. **103**, 064901 (2008).

[11]. G. Brenn, Droplet collision, in *Handbook of Atomization and Sprays* (Springer, Berlin, 2011).

[12]. H. P. Kavehpour, Coalescence of drops, Annu. Rev. Fluid Mech. **47**, 245 (2015).

[13]. M. Orme, Experiments on droplet collisions, bounce, coalescence and disruption, Prog. Energy Combust. Sci. **23**, 65 (1997).

[14]. K.-L. Pan, P.-C. Chou, and Y.-J. Tseng, Binary droplet collision at high Weber number, Phys. Rev. E **80**, 036301 (2009).

[15]. N. Roth, C. Rabe, B. Weigand, F. Feuillebois, and J. Malet, Droplet collision outcomes at high Weber number, in Proc. 21st Conf. *Institute for Liquid Atomization and Spray Systems* (ILASS, 2007).

[16]. G. Finotello, J. T. Padding, N. G. Deen, A. Jongsma, F. Innings, and J. Kuipers, Effect of viscosity on droplet-droplet collisional interaction, Phys. Fluids **29**, 067102 (2017).

[17]. C. Gotaas, P. Havelka, H. A. Jakobsen, H. F. Svendsen, M. Hase, N. Roth, and B. Weigand, Effect of viscosity on droplet-droplet collision outcome: Experimental study and numerical comparison, Phys. Fluids **19**, 102106 (2007).

[18]. C. Tang, P. Zhang, and C. K. Law, Bouncing, coalescence, and separation in head-on collision of unequal-size droplets, Phys. Fluids **24**, 022101 (2012).

[19]. C. Tang, J. Zhao, P. Zhang, C. K. Law, and Z. Huang, Dynamics of internal jets in the merging of two droplets of unequal sizes, J. Fluid Mech. **795**, 671 (2016).

[20]. D. Zhang, C. He, P. Zhang, and C. Tang, Mass interminglement and hypergolic ignition of TMEDA and WFNA droplets by off-center collision, Combust. Flame **197**, 276 (2018).

[21]. Z. Zhang, Y. Chi, L. Shang, P. Zhang, and Z. Zhao, On the role of droplet bouncing in modeling impinging sprays under elevated pressures, Int. J. Heat Mass Transf. **102**, 657 (2016).

[22]. Z. Zhang, and P. Zhang, Cross-Impingement and Combustion of Sprays in High-Pressure Chamber and Opposed-piston Compression Ignition Engine, Appl. Therm. Eng. (2018).

[23]. Z. Zhang, and P. Zhang, Modeling Kinetic Energy Dissipation of Bouncing Droplets for Lagrangian Simulation of Impinging Sprays under High Ambient Pressure, At. Sprays **28**, (2018).

[24]. P. O'Rourke, and F. Bracco, Modeling of drop interactions in thick sprays and a comparison with experiments, Proc. Inst. of Mech. Eng. **9**, 101 (1980).

[25]. S. L. Post, and J. Abraham, Modeling the outcome of drop–drop collisions in Diesel sprays, Int. J. Multiph. Flow **28**, 997 (2002).

[26]. R. Brown, and L. Scriven, The shape and stability of rotating liquid drops, Proc. R. Soc. Lond. A **371**, 331 (1980).

[27]. H. Kitahata, R. Tanaka, Y. Koyano, S. Matsumoto, K. Nishinari, T. Watanabe, K. Hasegawa, T. Kanagawa, A. Kaneko, and Y. Abe, Oscillation of a rotating levitated droplet: Analysis with a mechanical model, Physical Review E **92**, 062904 (2015).

[28]. J. Holgate, and M. Coppins, Shapes, stability, and hysteresis of rotating and charged axisymmetric drops in a vacuum, Phys. Fluids **30**, 064107 (2018).

[29]. E. K. Poon, J. Lou, S. Quan, and A. S. Ooi, Effects of streamwise rotation on the dynamics of a droplet, Phys. Fluids **24**, 082107 (2012).

[30]. B. Maneshian, K. Javadi, M. T. Rahni, R. Miller, Droplet dynamics in rotating flows, Adv. Colloid Interface Sci **236**, 63 (2016).

[31]. E. Janiaud, F. Elias, J. Bacri, V. Cabuil, and R. Perzynski, Spinning ferrofluid microscopic droplets, Magnetohydrodynamics **36**, 301 (2000).

[32]. S. Popinet, An accurate adaptive solver for surface-tension-driven interfacial flows, J. Comput. Phys. **228**, 5838 (2009).

[33]. S. Popinet, Numerical models of surface tension, Annu. Rev. Fluid Mech. **50**, 49 (2018).

[34]. X. Chen, D. Ma, P. Khare, and V. Yang, Energy and mass transfer during binary droplet collision, in 49th AIAA *Aerospace Sciences Meeting* (AIAA, 2011).

[35]. X. Chen, D. Ma, and V. Yang, Collision outcome and mass transfer of unequal-sized droplet collision, in 50th AIAA *Aerospace Sciences Meeting* (AIAA, 2012).

[36]. X. Chen, and V. Yang, Thickness-based adaptive mesh refinement methods for multi-phase flow simulations with thin regions, J. Comput. Phys. **269**, 22 (2014).

[37]. C. He, X. Xia, and P. Zhang, Non-monotonic viscous dissipation of bouncing droplets undergoing off-center collision, Phys. Fluids **31**, 052004 (2019).

[38]. C. Hu, S. Xia, C. Li, and G. Wu, Three-dimensional numerical investigation and modeling of binary alumina droplet collisions, Int. J. Heat Mass Transf. **113**, 569 (2017).

[39]. X. Xia, C. He, D. Yu, J. Zhao, and P. Zhang, Vortex-ring-induced internal mixing upon the coalescence of initially stationary droplets, Phys. Rev. Fluids **2**, 113607 (2017).

[40]. X. Xia, C. He, and P. Zhang, Universality in the viscous-to-inertial coalescence of liquid droplets, PNAS **116**, 23467 (2019).

[41]. P. Zhang, and C. K. Law, An analysis of head-on droplet collision with large deformation in gaseous medium, Phys. Fluids **23**, 042102 (2011).

[42]. E. Coyajee, and B. J. Boersma, Numerical simulation of drop impact on a liquid–liquid interface with a multiple marker front-capturing method, J. Comput. Phys. **228**, 4444 (2009).

[43]. C. He, X. Xia, and P. Zhang, Vortex-dynamical implications of nonmonotonic viscous dissipation of off-center droplet bouncing, Phys. Fluids **32**, 032004 (2020).